\documentclass[sigconf]{acmart}

\usepackage[colorinlistoftodos]{todonotes}
\usepackage{verbatim}
\usepackage{tabularx}
\usepackage{soul}
\usepackage{enumitem}
\usepackage{booktabs}
\usepackage{blindtext}
\usepackage{cleveref}
\usepackage{todonotes}

\usepackage{caption}
\usepackage{subcaption}

\graphicspath{ {figures/} }

\def\BibTeX{{\rm B\kern-.05em{\sc i\kern-.025em b}\kern-.08emT\kern-.1667em\lower.7ex\hbox{E}\kern-.125emX}}
    
\newcommand{\etal}{\emph{et al.}\xspace}
\newcommand{\change}[1]{{\color{black} #1}}
\begin{document}

%
\title{Comments on Comments: Where~Code~Review~and~Documentation~Meet}
%

\author{Nikitha Rao}
\email{nikitharao@cmu.edu}
\affiliation{%
  \institution{Carnegie Mellon University}
  \state{United States}
}

\author{Jason Tsay}
\email{jason.tsay@ibm.com}
\affiliation{%
  \institution{IBM Research}
  \state{United States}
}

\author{Martin Hirzel}
\email{hirzel@us.ibm.com}
\affiliation{%
  \institution{IBM Research}
  \state{United States}
}

\author{Vincent J. Hellendoorn}
\email{vhellendoorn@cmu.edu}
\affiliation{%
  \institution{Carnegie Mellon University}
  \state{United States}
}

%

\begin{abstract}
A central function of code review is to increase understanding; helping reviewers understand a code change aids in knowledge transfer and finding bugs. Comments in code largely serve a similar purpose, helping future readers understand the program. 
It is thus natural to study what happens when these two forms of understanding collide. We ask: what documentation-related comments do reviewers make and how do they affect understanding of the contribution?
We analyze ca.~700K review comments on 2,000 (Java and Python) GitHub projects, and propose several filters to identify which comments are likely to be either in response to a change in documentation and/or call for such a change. We identify 65K such cases.
We next develop a taxonomy of the reviewer intents behind such ``comments on comments''. We find that achieving a shared understanding of the code is key: reviewer comments most often focused on clarification, followed by pointing out issues to fix, such as typos and outdated comments. Curiously, clarifying comments were frequently suggested (often verbatim) by the reviewer, indicating a desire to persist their understanding acquired during code review.
We conclude with a discussion of implications of our comments-on-comments dataset for research on improving code review, including the potential benefits for automating code review.

\end{abstract}


%
\maketitle
\section{Introduction}


Code review is an important mechanism for preventing bugs in software projects \cite{kemerer2009}. To do so effectively, reviewers focus heavily on understanding the code and change made to it \cite{bacchelli2013expectations}. Natural language communication is a key component here: reviewers use comments to discuss new ideas and to overcome issues \cite{tsay2014let}, as well as transfer knowledge, social norms and conventions \cite{dabbish2012}.

Another form of natural language associated with programming is documentation, in the form of block or in-line comments in the code. Studies have shown that these help developers better understand code~\cite{tenny88, woodfield1981}.
Pascarella \etal~\cite{pascarella17} manually annotated 2,000 Java files to develop a taxonomy of code comment types and found that `summary' and `usage' were the most prominent categories implying that comments mainly serve as a means for transferring knowledge, especially to end-user developers. Aghajani \etal empirically studied 878 documentation-related artifacts and found that issues related to the correctness of code comments, such as insufficient or obsolete content, seemed most common \cite{aghajani19}. A survey confirmed that these issues negatively affected understanding and made it more challenging to use others' code \cite{aghajani20}.


Given that code review heavily relies on natural language to facilitate understanding and that code comments serve much the same purpose, we wonder: do contributors and project maintainers give consideration to documentation during code review, and, if so, how does documentation factor into the review?
To the best of our knowledge, no prior work has studied this coalescence of ``comments on comments''; thus, in this work we provide a first, data-driven exploration at the intersection of these two domains.

\begin{figure*}[ht]
    \centering
    \includegraphics[width=.9\linewidth]{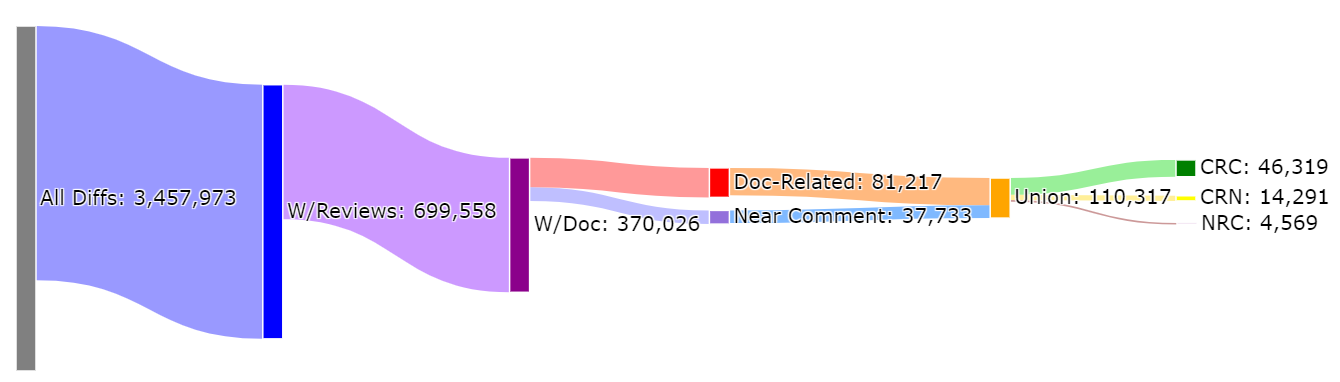}
    \vspace{-3mm}
    \caption{Breakdown of the data collection process leading to the three categories of comments-on-comments that we identify (see \Cref{sec:dataset}). 
    Note that the scaling in the first transition has been altered to better emphasize the subsequent distributions.}
    \vspace{-4mm}
    \label{fig:data_split}
\end{figure*}

\begin{figure*}[t]
    \centering
    \includegraphics[width=.82\linewidth]{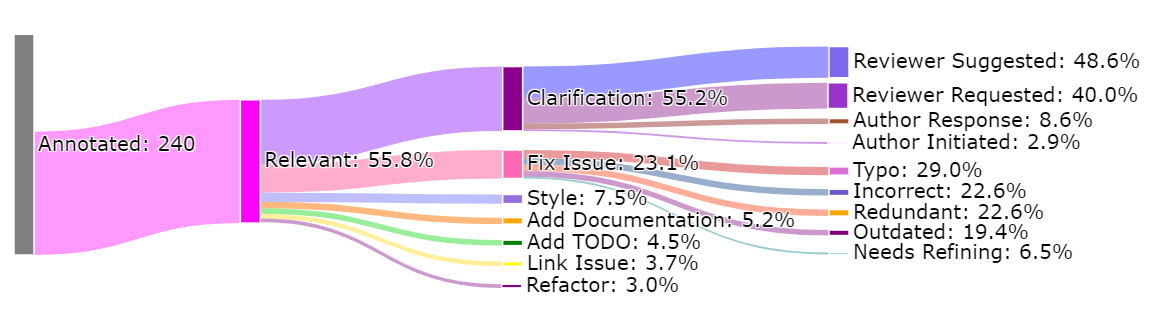}
    \caption{Taxonomy of review comment intents, based on 80 samples from each of the three comments-on-comments categories. Ca.~56\% of these were indeed code-comment related discussions, of which more than half related to clarification.}
    \label{fig:annotations}
\end{figure*}

Specifically, we commence by quantitatively analyzing how often reviewers pay attention to changes made, and/or request that changes be made, to documentation in contributed code.
We approach this by mining code review data from the event streams of 1K popular GitHub repositories in Python and Java each, with 700K review comments between them.
We filtered this data to identify cases that likely represent documentation-related review comments  (\Cref{fig:data_split}). We identify three salient types of comments on comments, ordered by frequency:
\begin{itemize}[nosep]
    \item[(CRC)] A revision includes a change to a code comment (C); the reviewer comments on it (R), following which the developer updates the code comment again (C).
    \item[(CRN)] A revision included a change to a code comment (C), the reviewer comments on it (R), but this does not result in a further change (N).
    \item[(NRC)] A revision did not change (N) a code comment (only the code was modified), then following a review comment (R) the developer updates a code comment (C).
\end{itemize}
We next conduct a qualitative analysis of samples from each of these categories to derive a taxonomy of \emph{intents} behind reviewer comments (\Cref{fig:annotations}). We find that reviewers frequently ask for clarification related to the code or comment change, indicating that the initial comment revision (or lack thereof) inadequately captured the change. Interestingly, they often state the clarification required, regularly verbatim. This highlights a desire for reviewers to \emph{ensure that their understanding of the code (change) is persisted,} for which in-code documentation is a natural fit.

Our findings show that code comments are an important consideration in code review in open-source projects, and that such comments frequently lead to improved documentation. Besides advancing our understanding of code review, we expect these findings to support efforts towards improving and automating code review, where AI-based models \cite{tufano2021} have struggled to capture the full scope of code comments seen in practice \cite{hellendoorn2021towards}. Incorporating information from (changes to) code comments may help these models better understand reviewing needs. The CommentsOnComments dataset can be found here - \url{https://doi.org/10.5281/zenodo.5886145}

\section{Methodology}
\label{sec:methodology}
We collect our data from events stored on the GitHub Archive.\footnote{https://www.gharchive.org/} We focus on events related to pull request (PRs) belonging to the 1,000 projects with the most review comments in Java and Python, archived between January 2017 to November 2020. We split each pull request into a series of ``diff chunks'' that modify successive line(s), resulting in \textasciitilde{}3.5 million such diffs (ca. 2M from Java and 1.4M from Python). Following the process outlined by Hellendoorn \etal~\cite{hellendoorn2021towards}, we then align each code review comment, which comes with an associated diff at the time of said comment (as in \Cref{fig:examples}), with one of the diff chunks in the final PR. This gives us a triple of $\langle$initial change, review comment, final change$\rangle$. Out of the 3.5M total diff chunks, ca. 700K were aligned with a review comment in this way.\footnote{In fairly rare cases, a diff chunk may be commented on multiple times.}

\subsection{Data Pre-Processing}
\label{sec:preprocessing}
We next apply a series of filters to identify review comments that are likely documentation-related -- we call these ``comments on comments''. Following the flow in \Cref{fig:data_split}, we first identify diff chunks with associated review comments (\textasciitilde{}700K), and next subset diffs containing changes to in-line or block code comments (either before or after the reviewer's comments, \textasciitilde{}370K). We found that the presence of both types of comments alone is rarely indicative of a code comment-related review comment (diffs can be quite large), so we apply two filters to find reviewer comments that are likely to be documentation-related:

\noindent\textbf{Documentation-related comment.} This filter identifies cases where reviewers explicitly mention documentation. We look for both the following documentation related keywords: \{\textit{documentation, docstring, javadoc, comment, todo}\} \ul{and} for review comments containing code suggestions (a commonly used feature on GitHub, see \Cref{fig:ex3}) that involve changes to inline comments (i.e., containing one of \textit{ //, /*, */} for Java and \textit{\#, """, `{}`{}`} for Python).
We constructed the list of keywords based on a manual analysis of random samples from the aforementioned 370K subset in which we iteratively added keywords that were used to refer to code comments until no new words appear. \change{To elaborate, we started with the keywords “comment” and “documentation”, from which one annotator used snowballing to construct the final list across 3 iterations with 60 diff chunks each (not included in 240 final samples used in Section~\ref{sec:taxonomy}). The annotator added new keywords that predominantly occurred in reviewer comments on documentation-related samples identified via the initial set of keywords.} This occasionally yields false positives; e.g., reviewers sometimes use the keyword `comment' to refer to another/previous review comments (e.g., \textit{`Please refer to my previous comment}').

\noindent\textbf{Comment near comment.} We expect that a reviewer comment placed `near' a code comment is likely to relate to it. Due to GitHub's UI design, which typically shows three or more lines preceding the line being commented on, reviewer comments often refer to lines of code placed above them within that window. Thus, we mark a review comment as `near' a code comment if the latter is present in that window. This is especially useful in filtering out false positives in large diff chunks with multiple review comments.

As \Cref{fig:data_split} shows, the first filter yields the most cases (81K). These filters are also fairly disjoint, yielding 110K samples in their union, thus reducing the naive alignment by around 70\%. While Java had more documentation related comments, Python had more comments near comments, so both languages contribute roughly evenly to this filtered set (ca. 60K and 50K respectively).

\change{\noindent\textbf{Evaluation of filters.} We compute the recall and precision scores for the filters used using the labels obtained from the manual annotation performed in Section \ref{sec:taxonomy}. The documentation-related comments filter has a recall of 79.7\% and precision of 66.7\% whereas the comments-near-comments filter has a recall of 46.7\% and a precision of 43.9\%. Note that these scores do not impact our findings in RQ2 and RQ3.}

\subsection{Comments On Comments Dataset}
\label{sec:dataset}
We categorize the resulting documentation-related review comments based on the preceding and subsequent changes to the corresponding documentation. We consider three cases: two in which the review comments are likely to be a response to a change in documentation (CRC \& CRN, depending on what happens next), and one in which it likely called for such a change (NRC). To elaborate:

\textbf{CRC} - Changed code comment (C) followed by review comment~(R), and then another change to the code comment (C). 

\textbf{CRN} - Changed code comment (C), then review comment (R), no further changed code comment (N). The reviewer comment lead to no change at all or the changes were directed only towards the code.

\textbf{NRC} - In this case, the initial revision affected some code but not (N) its nearby comment. The reviewer comment (R) then triggers a further change, this time involving the nearby code comment (C).

Note that we omit a fourth scenario in which no change was made to a code comment both before and after a review comment, since we found reviews in such cases to be rarely related to documentation in a preliminary analysis. 
As shown in \Cref{fig:data_split}, these categories comprise the majority of the 110K samples filtered earlier, with the CRC category the most common.

\subsection{Qualitative Analysis \& Taxonomy}
\label{sec:taxonomy}
\change{Based on a casual inspection of 30 randomly chosen samples from the comments on comments data (10 per category), we observed that reviewers talked about documentation in several different ways.} To derive a more formal taxonomy of reviewer intents, we randomly choose 40 samples from each of the three cases (CRC, CRN and NRC) for both Java and Python and distribute the resulting 240 samples across three annotators.
For each datapoint, the annotator was required to look at both the review comment and the diff chunk before and after the review comment was placed, and note the following: (i)~whether the review comment was documentation-\emph{relevant}; i.e. it directly discussed documentation and/or triggered a change to documentation, (ii)~the high-level reviewer \emph{intent} for the review comment, and (iii)~any additional notes capturing finer details or sub-intents. \change{Additionally, annotators were asked to make a note of any ambiguous samples.}
Note that each annotator looked at a distinct subset due to the lack of ambiguity in this task.
For (ii), the goal was to summarize the purpose of the comment in one or a few words; annotators were not constrained to a fixed vocabulary, but were given several examples, such as ``clarification'' and ``fix issue'', based on a preliminary annotation for reference. \change{Note that the reference captured common trends seen during the exploratory analysis. While this reference included several examples of reviewer comments with different intents, it did not serve as an exhaustive set of intents for annotators to consider -- several categories including ``Add TODO'' emerged during annotation.}
We observed that there was a significant overlap in the intent categories derived by each annotator, with an exception of a few rare corner cases. \change{Less than 3\% (7/240) were marked ambiguous; these were resolved by discussion with all three annotators with 100\% negotiated agreement (2/7 samples were assigned multiple intents).} This was followed by a discussion among the annotators which resulted in a coordinated taxonomy of intent categories and (for the largest categories) sub-intents. 
We did not observe any noticeable differences in the intent categories found for Java and Python.
\Cref{fig:annotations} summarizes the distribution of intents in the annotated data. 
\section{Results}
\label{sec:results}
We seek to answer three research questions using our data set. \\

\noindent\textbf{RQ1: How often do review comments concern code comments?}\\
We ground our discussion in an analysis of the prevalence and characteristics of code comment related review comments. 
First, out of the 700K diffs that receive review comments, we identify ca. 370K that contain code comments. It is worth noting that only ca. 29\% (\textasciitilde{1M} out of 3.5M) of the initial set of diffs (commented on or not) contained documentation, implying that such diffs are relatively more likely to be commented on (370K/1M = 37\% vs. 330K/2.4M = 14\%).\footnote{This need not be a causal relationship; more complex code may simply require both more documentation and reviewer attention.}
Based on the filters applied in \Cref{sec:methodology}, we find that 15.8\% (110K/700K) of the diffs with reviewer comments are plausibly being commented on in relation to documentation. Breaking this down further, 11.6\% (81K) of the reviewer comments explicitly talked about documentation and 5.4\% (38K) of the reviewer comments are near code comments (leaving a 0.8\% overlap between these cases).

We further dissected these 110K samples based on the three scenarios described in \Cref{sec:dataset}. We find that reviewers are much more likely to comment on a diff chunk when the contributor had already made some change to a code comment (50.8\%) compared to diff chunks that only made changes to the code (15.8\%). \change{Extrapolating from the manual annotation performed in RQ3, we find that our estimate of 55.8\% of samples being documentation-\textit{relevant}, i.e. the comment discussed documentation and/or triggered a change to documentation, comes with a 95\% confidence interval of ~6.5\% using the Clopper-Pearson exact method. This indicates that between 32,133 and 40,554 of the complete set of samples is likely to be documentation-\textit{relevant}.}

\textit{Finding: Reviewers frequently consider documentation when reviewing code. They are more likely to comment when contributors updated code comments initially.}\\

\noindent\textbf{RQ2: Do reviewer comments affect the documentation of contributed code?}\\
Within the 65K ``comments on comments'' cases, where a review comment is either in response to a change in code comment or a call for such a change, we find that a review comment in response to a change made to the code comment lead to the code comment getting updated (CRC) 76.4\% (46K/61K) of the time, with less than a quarter of reviewer comments resulting in no further change to the code comments (CRN). In comparison, review comments on code changes that previously did not change a comment only resulted in the code comment getting updated (NRC) in 26.8\% of cases. We also observed that reviewer comments were more likely to lead to a comment getting updated in Python compared to Java. \change{ More specifically, following an initial change in documentation and review comment (CRC vs. CRN), the odds ratio of a further change in documentation in Python vs. Java is \mbox{2.02 ($\pm$0.08,} 95\% confidence interval). In contrast, the odds ratio irrespective of an initial documentation change (CRC+NRC vs. CRN) is \mbox{1.25 ($\pm$0.05,} 95\% confidence interval).}


\textit{Finding: Review comments in response to a code comment change often result in the code comment getting updated.}\\

\begin{figure}
     \centering
     \begin{subfigure}[b]{0.48\textwidth}
         \centering
         \includegraphics[width=\textwidth]{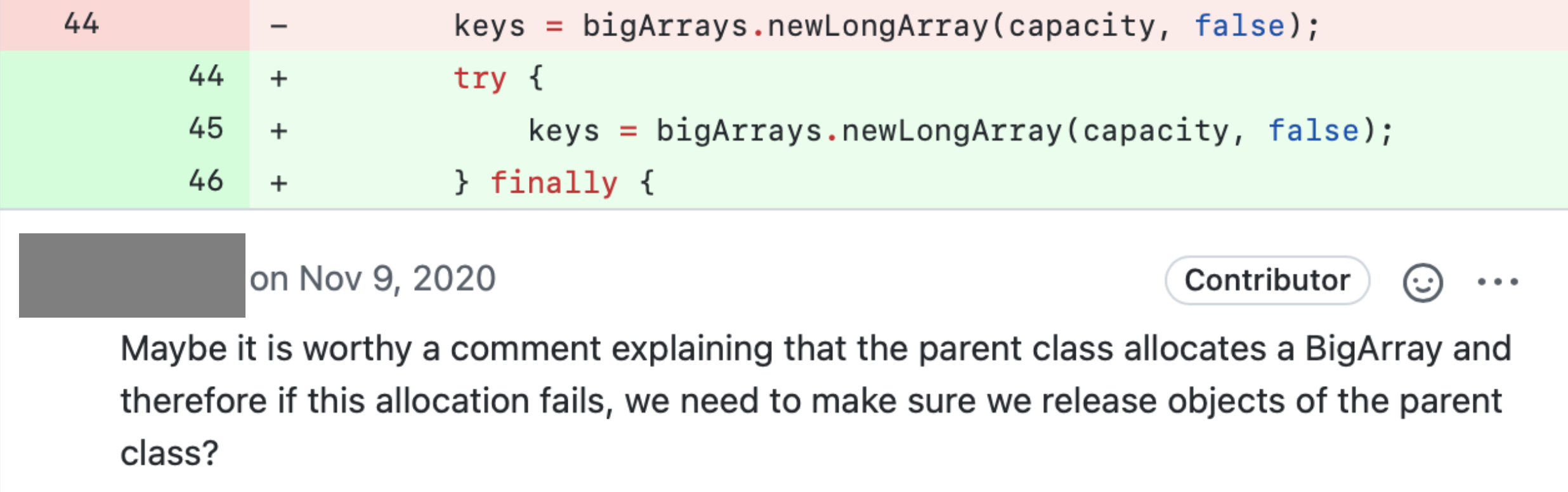}
        \vspace{-6mm}
         \caption{NRC + intents: clarification - reviewer suggested \\ (github.com/elastic/elasticsearch/pull/64744)}
         \label{fig:ex1}
     \end{subfigure}
     \hfill
     \begin{subfigure}[b]{0.48\textwidth}
         \centering
         \includegraphics[width=\textwidth]{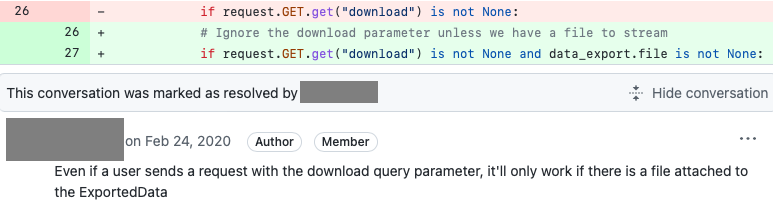}
        \vspace{-6mm}
         \caption{CRC + intents: clarification - author initiated \\ (github.com/getsentry/sentry/pull/17267)}
         \label{fig:ex2}
     \end{subfigure}
     \hfill
     \begin{subfigure}[b]{0.48\textwidth}
         \centering
         \includegraphics[width=\textwidth]{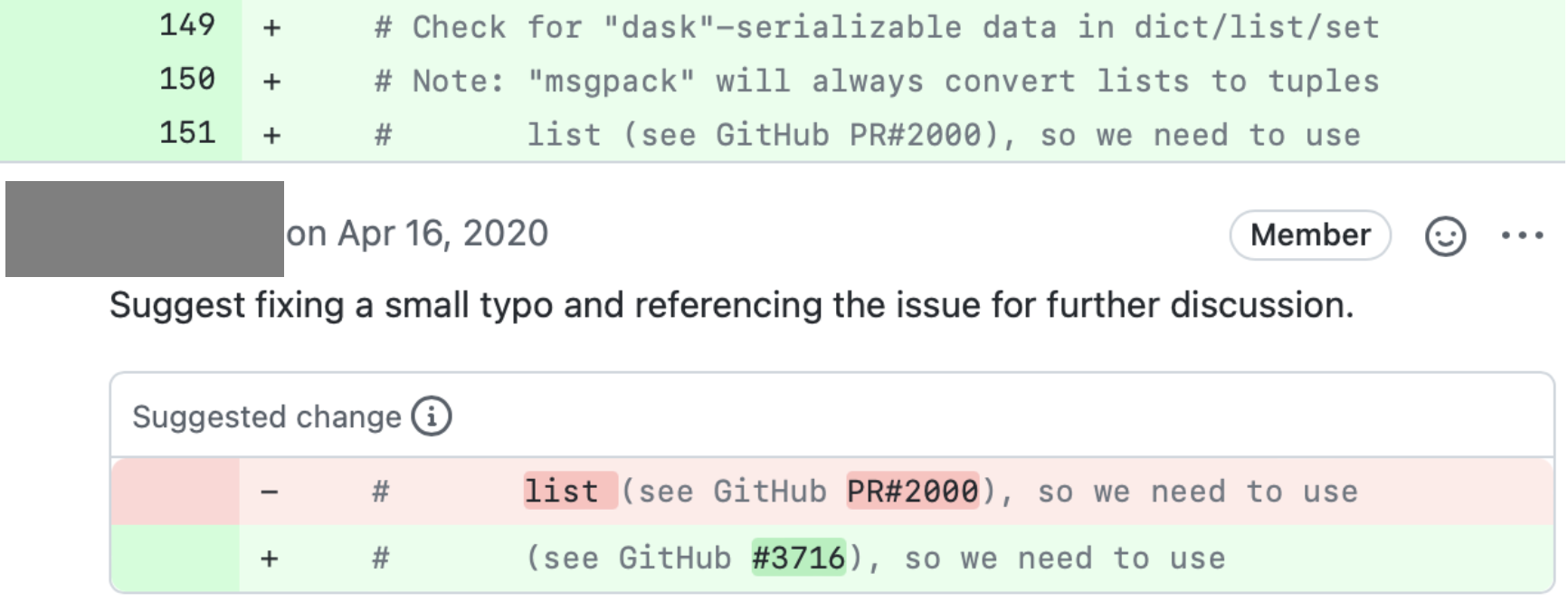}
        \vspace{-6mm}
         \caption{CRC case, intents: fix issue - typo, link issue \\ (github.com/dask/distributed/pull/3689)}
         \label{fig:ex3}
     \end{subfigure}
     \vspace{-8mm}
        \caption{Examples of comments on comments with annotations.}
        \vspace{-2mm}
        \label{fig:examples}
\end{figure}

\noindent\textbf{RQ3: What are the intents of reviewing documentation?}\\
We next conduct a qualitative analysis of review comments from each category based on our manual annotation following the protocol described in \Cref{sec:taxonomy}. This results in the taxonomy of reviewer intents for changes made to code comments that is visualized in \Cref{fig:annotations}. We found that \emph{clarification} was the most common goal (70/134 documentation-\textit{relevant} comments) of a review comment, where some change needed to be made to the documentation to help understand the updated code. This type of change was often initiated by the reviewer, either as a \emph{request} (28/70) to elaborate on a change, or a (regularly verbatim) \emph{suggestion} (34/70) of what to comment to help clarify the new code for others. \Cref{fig:ex1} shows such a case, in which a reviewer encourages the author to document the need for the added \textit{try/catch} block by prescribing what that comment should say. Less commonly, the author would either \emph{respond} (6/70) with clarification to a comment or code change suggested by the reviewer or preemptively \emph{initiate} (2/70) a discussion related to the change. \Cref{fig:ex2} illustrates the latter case, in which an author reviewed their own code after submission and later added an informative code comment reflecting the one seen here.

The next-most common intent was \emph{fix issue} (31/134), where reviewers mostly asked to \emph{fix typos} (10/31), update comments that were \emph{inaccurate/ambiguous} (7/31), remove comments that were \emph{redundant} (7/31), or update newly-\emph{outdated} comments (6/31). We also identified several other, less common intent categories, including where reviewers asked to make a \emph{style} related change (10/134, e.g. adjusting capitalization), or to add either a \emph{Javadoc/Docstring} (7/134), a \emph{to-do} statement (6/134), or a link to the corresponding GitHub \emph{issue} (5/134), or requested a \emph{refactoring} (4/134) such as moving the comment to a different location.
\Cref{fig:ex3} shows an example of both a typo fix and issue link addition being recommended through GitHub's ``suggested change'' option. \change{We found little difference in intent distributions between Python and Java. For example, the clarification intent prevalence was 52.7\% in Python and 51.6\% in Java; the ``fix issue'' ratios were 25\% and 20\% respectively. The largest difference was in the relevance of samples for the CRN case (67.5\% for Python and 35\% for Java).}

\textit{Finding: Most reviewer comments seek some form of clarification to increase or enshrine their understanding of the change made to the code, followed by fixing issues in the comments themselves.}

\section{Discussion}
\label{sec:discussion}

Our work at the intersection of code review and documentation highlights that achieving a shared understanding of code is vital in open source project maintenance. Previous work established that code review is a key mechanism for transferring knowledge and reviewers spend much of their time on understanding a contribution~\cite{bacchelli2013expectations} for which they rely on discussions and during which they impart their knowledge of the project to the contributor in turn~\cite{dabbish2012}. The results from our analysis suggest that when reviewers and authors discuss documentation, it is mostly to clarify the code or documentation changes. We observe that this facilitates understanding not only for reviewers but also often for contributors.

Code review discussions often explicitly or implicitly involve third party stakeholders in the ``audience''~\cite{tsay2014let}. Our results clearly show that discussing documentation not only helps to achieve a shared understanding between reviewers and authors, but also benefits the understanding of other, future readers. For instance, reviewers frequently demonstrated that they personally understood a change, but also suggested documentation changes to ensure that future readers would share that understanding. Our dataset is well-suited towards better understanding the nuances behind this understanding process by aligning actual changes in code comments to related review discussions. For example, 74.3\% of the clarification discussions mentioned above resulted in subsequent revisions to the relevant documentation. This suggests at least some success in achieving a shared understanding of the code.

Our results also show that reviewers are spending considerable attention and effort on documentation, as evidenced by their higher response rate when documentation is updated in a revision and by the many cases of review comments triggering documentation changes. Given that achieving understanding of a code contribution is particularly time consuming for reviewers~\cite{bacchelli2013expectations}, and that recent efforts have shown significant challenges in automating code review using AI \cite{tufano2021,hellendoorn2021towards}, our findings may offer a path forward for supporting tool-based understanding of code changes. Namely, such tools should both earnestly consider relevant documentation while also expecting it to be often at least somewhat flawed. And they should perhaps take a cue from real-life reviewers and learn to ask for (or suggest) clarification. Such dialogue-like agents are scarce in software engineering practice, but our results hint at the potential importance of this ability.

\paragraph{Threats to Validity}
We designed several filters to identify 65K comments that are plausibly documentation-related. Some of these filters are conservative: the list of keywords used to identify docu\-mentation-related comments was fairly short, so our analysis may miss some cases. Vice versa, our manual annotation showed that these filters also include around 44\% non-documentation relevant comments, which may impact the absolute tallies in our quantitative analysis. Encouragingly, we did not find that any language or comments on comments category was significantly less likely to be relevant than others, so we expect our relative trends to hold.
Our taxonomy was derived based on an even sampling of the three types of comments-on-comments, so the distributions of intents in the taxonomy may not be representative of those in the wild. This was an intentional decision, as we wanted the taxonomy to cover a broad span of motivations. While sample sizes are too small to run statistical tests, we did not notice a strong bias of intents towards certain cases.\footnote{We did notice a slight, and intuitively plausible, preference for clarification comments to originate from reviewers unprompted by a comment change (NRC) and for ``fix issue'' comments to occur in the CRC case. Future work may study this effect in more detail.} Lastly, we limit our study to only changes made to inline comments (and not block comments unless they have a documentation related reviewer comment associated with it) for the 3 cases.
\section{Conclusion}
\label{sec:conclusion}
We present the first data-driven study at the intersection of code review comments and in-code documentation comments. We find these to be a relatively common topic of discussion during code reviews among a large number of GitHub projects, in particular when the initially submitted changeset affected documentation comments to begin with. Review comments in such locations frequently spurred further revisions to documentation, most commonly to add clarifications for the benefit of future readers, highlighting that reviewers are concerned with persisting their understanding of code changes. Our quantitative and qualitative analysis highlights the important role of documentation in the discussion of code contributions. We release our data to benefit further studies.
%
\bibliographystyle{ACM-Reference-Format}
\bibliography{references}

\end{document}